Transport properties of the single- and 3-core Fe-Se wires fabricated by a novel chemical-transformation PIT process

Yoshikazu Mizuguchi[1], Hiroki Izawa[1], Toshinori Ozaki[2], Yoshihiko Takano[2] and Osuke Miura[1]

1. *Department of Electric and Electronic Engineering, Tokyo Metropolitan University, 1-1 Minami-osawa, Hachioji 192-0397, Japan*

2. *National Institute for Materials Science, 1-2-1 Sengen, Tsukuba 305-0047, Japan*

Abstract

We fabricated single- and 3-core superconducting Fe-Se wires using a novel process based on a chemical transformation from hexagonal $FeSe_{1+d}$ (non-superconducting) to tetragonal FeSe (superconducting) via an optimal supply of Fe from the Fe sheath by annealing. This process enhanced a packing density of superconducting core inside the sheath, owing to an expansion of the lattice volume via a chemical transformation from high-density hexagonal $FeSe_{1+d}$ to low-density tetragonal FeSe. The obtained superconducting wire showed superconductivity below ~10 K and upper critical field was estimated to be 19.3 T at 0 K. The obtained transport critical current densities at 0 T were 588 A/cm² for the 3-core wire and 218 A/cm² for the single-core wire.

The discovery of Fe-based superconductors have activated studies on its applications as thin films and superconducting wires owing to the high transition temperature ($T_c$) and the high upper critical field ($H_{c2}$) [2-4]. Among the Fe-based superconductors, FeSe is one of the materials that should be mostly expected to realize practical applications, because of the simple binary composition, high $H_{c2}$, and low anisotropy [5]. Furthermore, the $T_c$ reaches 37 K under high pressure around 4-6 GPa, while the ambient $T_c$ is only ~10 K [6-9]. Large enhancements of $T_c$ have been observed in strain-stressed thin films, bulk poly crystals, and wires [10-12]. In these respects, FeSe is a potential candidate for application such as superconducting wires and thin films, particularly under high magnetic fields.

Requirements for achieving a high critical current density ($J_c$) in a superconducting wire are high quality of superconducting core (superconducting material in the sheath), good connectivity between the sheath and superconducting core, and high packing density of the superconducting core. To date, fabrications of Fe-chalcogenide superconducting wires have been based on the in-situ powder-in-tube (PIT) method or the ex-situ PIT method with an Fe sheath, in which chalcogen powders (for in-situ) or Fe-chalcogenide powders (for ex-situ) were used as the precursor and packed into a Fe tube [12-16]. Although the in-situ PIT method produced better connections between the sheath and the superconducting core owing to a direct chemical reaction between the sheath and precursor, some pores arising from evaporation of chalcogen and/or drastic diffusion of inner elements during the reaction were observed; hence the packing density was low. Although, by using the ex-situ PIT method, comparably high packing density was obtained due to almost no chemical reaction during wire fabrication process, high $J_c$ was not observed, probably due to bad connections between the sheath and the core. In these respects, new fabrication process which can realize both a higher packing density and good connectivity between the sheath and the superconducting core is required.

Here we report transport properties of single- and 3-core superconducting FeSe wires fabricated by a novel chemical-transformation PIT process based on a chemical transformation from hexagonal $FeSe_{1+d}$ (non-superconducting) to tetragonal FeSe (superconducting) via an optimal supply of Fe from the Fe sheath by annealing in the wire fabrication process, which is categorized as an intermediate between the in-situ and ex-situ method.

Poly crystals of $FeSe_{1.2}$ were prepared by a conventional solid state reaction method. Fe powders (99.9%) and Se tablets (99.999%) with a nominal composition of Fe:Se = 1:1.2 were sealed into an evacuated quartz tube, and heated at 700 ºC for 10

h. The obtained FeSe$_{1.2}$ powders were packed into an Fe tube with an outer diameter of 6.2 mm and inner diameter of 4.0 mm. The tube was sealed using Fe caps and groove-rolled into a rectangular rod wire with a size of ~2 mm for the single-core wire. For the 3-core wire, a rectangular rod with a size of ~1.6 mm was fabricated as same as single-core wire, and cut into several pieces with a length of 3 cm. These three wires were packed into the Fe tube, sealed with the Fe caps, and groove-rolled into a rectangular rod wire with a size of ~2 mm. The obtained single- and 3-core wires were cut into several pieces with a length of ~5 cm, sealed into an evacuated quartz tube, and annealed at 1000 ºC for 5 h.

The crystal structure was characterized by x-ray diffraction (XRD) using a CuKa radiation. Figure 1(a) shows the XRD patterns for the FeSe$_{1.2}$ precursor (before annealing). All the peaks were well indexed using a space group of *P*6$_3$/*mmc*, corresponding to the NiAs-type (hexagonal) structure depicted as an inset. The lattice constants were estimated to be *a* = 3.602(2) Å and *c* = 5.894(6) Å. Figure 1(b) shows the XRD profile for the powder inside the annealed wire (after annealing). Almost of the observed peaks were indexed using a space group of *P*4/*nmm*, corresponding to the PbO-type (tetragonal) structure, while the small peaks of pure Fe was also observed. The lattice constants of the tetragonal phase were estimated to be $a = 3.777(2)$ Å and $c = 5.535(7)$ Å. These results indicated that the chemical transformation from hexagonal (NiAs-type) FeSe$_{1+d}$ to tetragonal (PbO-type) FeSe was surprisingly achieved via a supply of Fe from the Fe sheath by short-time annealing for only 5 h. Figure 1(c) is a schematic diagram of the chemical-transformation PIT process. High packing density of the superconducting core can be obtained using this process, because the lattice volume of hexagonal FeSe$_{1+d}$ is smaller than that of tetragonal FeSe.

The cross section of wires was observed using an optical microscope. Figures 2(a) and (b) show optical microscope images of the cross section for the obtained single- and 3-core FeSe wires. It is clear that the FeSe core is well-packed in the Fe sheath and the boundary connectivity seems to be good. The cross-sectional area of the superconducting core for the single- and 3-core FeSe wires was estimated to be 0.0155 and 0.00517 cm$^2$, respectively.

Temperature dependence of resistance was measured using the four-terminal method under the magnetic field up to 7 T. Figure 3(a) displays the temperature dependence of resistance from 300 down to 2 K for the obtained single-core FeSe wire. The zero-resistance state was observed below 10 K, which is relatively higher than the $T_c$ of polycrystalline samples. The enhancement of $T_c$ might be due to a strain stress generated by the chemical transformation in the Fe sheath. Figure 3(b) shows the temperature dependence of resistance below 20 K under the magnetic fields of 0, 1, 3, 5 and 7 T. To obtain a magnetic field-temperature phase diagram, we estimated $T_c^{onset}$ and $T_c^{zero}$ using resistance criteria of 90 % and 10 % of the normal-state resistance just above $T_c$. Both the applied magnetic field and the estimated $T_c^{onset}$ and $T_c^{zero}$ were plotted in Fig. 3(c). From the linear extrapolations of upper critical field ($B_{c2}$) and irreversible field ($B_{irr}$), the $B_{c2}$ (0 K) and $B_{irr}$ (0 K) were estimated to be 28 and 21 T. By applying the WHH theory [17], which gives $B_{c2}$ (0 K) = $-0.69 T_c (dB_{c2}/dT)|_{T_c}$, the $B_{c2}$ (0 K) was estimated to be 19.3 T.

Current dependence of voltage was measured using the four-terminal method at 4.2 K. The length between the voltage taps were 8.2 and 7.8 mm for the single- and 3-core wires, respectively. Figure 4(a) shows the magnetic field dependence of $J_c$ for the single- and 3-core FeSe wires. The critical currents ($I_c$) were defined using a criterion of 1 μV/cm as indicated in Fig. 4(b, c), which shows the current dependences of voltage at 0 T for the single- and 3-core wires. The obtained $J_c$ (0 T) were 218 A/cm² for the single-core wire and 588 A/cm² for the 3-core wire. The $J_c$ of the 3-core wire was 2.7 times as large as that of the single-core wire, indicating that multi-core fabrication is effective to obtain high-$J_c$ FeSe superconducting wire using this method. This tendency implies that the $J_c$ near the boundary between the sheath and the FeSe core is higher than that of center of the core. Therefore, fabrication of the multi-core FeSe wire using the chemical-transformation PIT process is one of the ways to realize higher $J_c$ in FeSe superconducting wires. Another strategy is an enhancement of grain connectivity around the center of the FeSe core. As reported in FeAs-122 wires, addition of metals, such as Ag doping which improved the grain connectivity [18,19], might greatly enhance $J_c$ for FeSe wire as well. Furthermore, introduction of pinning centers are required to realize high $J_c$ for the next step.

In conclusion, we obtained single- and 3-core superconducting FeSe wires by a novel PIT process based on a chemical transformation from hexagonal $FeSe_{1+d}$ (non-superconducting) to tetragonal FeSe (superconducting) via an optimal supply of Fe from the Fe sheath into the core by annealing. The obtained superconducting wire possessed well-packed superconducting core, and the $T_c^{zero}$ and $H_{c2}$ were estimated to be ~10 K and 19.3 T, respectively. The obtained transport critical current densities at 0 T were 588 A/cm$^2$ for the 3-core wire and 218 A/cm$^2$ for the single-core wire. The higher $J_c$ in the 3-core wire suggests that multi-core wire is very effective to achieve higher $J_c$ in FeSe wires. The novel chemical-transformation process which we introduced here will be a great technique to fabricate high-$J_c$ Fe-based superconducting wires.

Acknowledgement

This work was partly supported by Grant-in-Aid for Scientific Research (KAKENHI). The crystal structures were depicted using VESTA [20].

Figure captions

Fig. 1. (a) XRD pattern for the FeSe$_{1.2}$ precursor. (b) XRD pattern for the powder collected from the obtained single-core wire. In figs. 1(a) and (b), the peaks for hexagonal and tetragonal phases are indicated using the symbols of " H " and " T ", respectively. The numbers in these figures are the mirror indices. (c) Schematic of the chemical-transformation PIT process. The crystal structure of the core transforms from the hexagonal NiAs structure (FeSe$_{1+d}$) to the tetragonal PbO structure (FeSe) via the wire fabrication process. Small (blue) and large (yellow) balls indicate Fe and Se ions, respectively.

Fig. 2. (a) Optical microscope image of the cross section for the single-core FeSe wire. (b) Optical microscope image of the cross section for the 3-core FeSe wire.

Fig. 3. (a) Temperature dependence of resistance from 300 to 2 K for the obtained single-core FeSe wire. (b) Temperature dependence of resistance below 20 K for the obtained single-core FeSe wire under magnetic fields up to 7 T. (c)Magnetic field-temperature phase diagram for the obtained single-core FeSe wire.

Fig. 4 (a) Magnetic field dependence of critical current density for the obtained single- and 3-core FeSe wires. (b,c) Current dependence of voltage for the obtained single-core and 3-core FeSe wires under 0 T. The lines in these figures indicate criteria of 1 μV/cm to estimate $I_c$.

Fig. 1

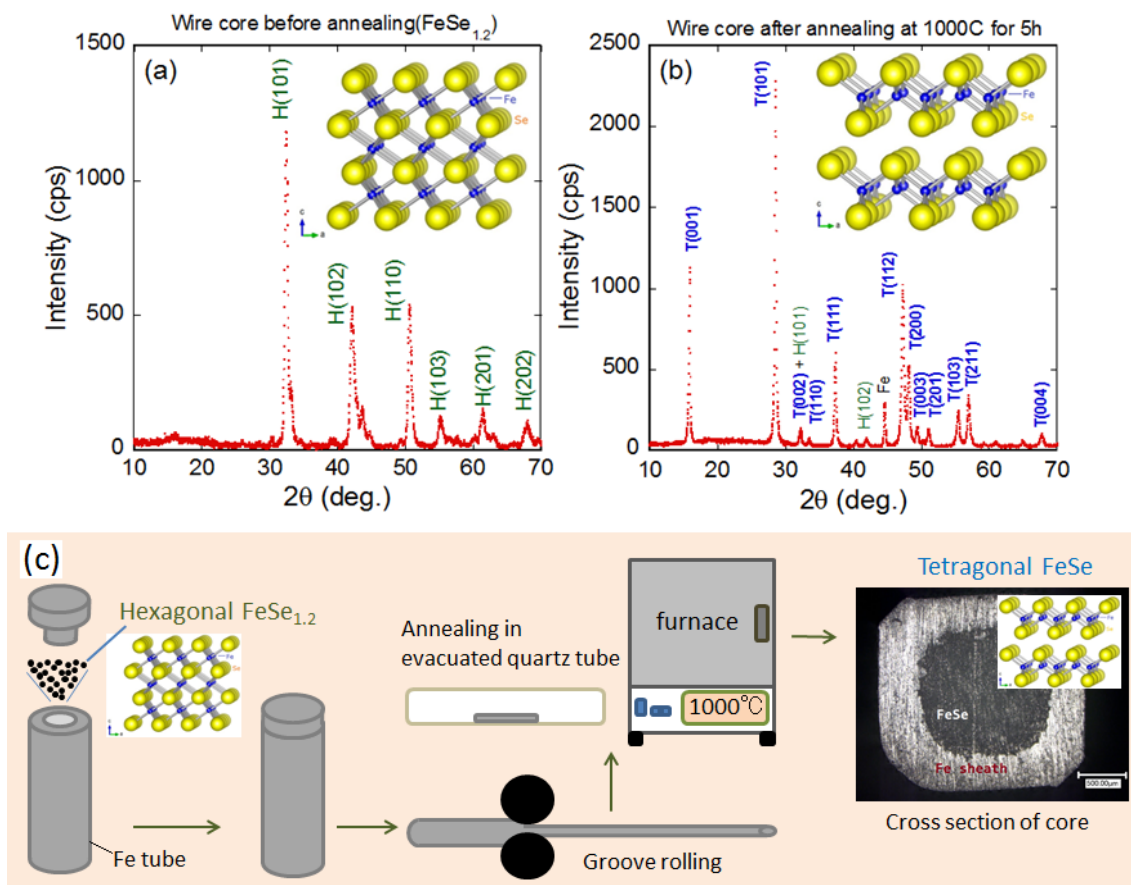

Fig. 2

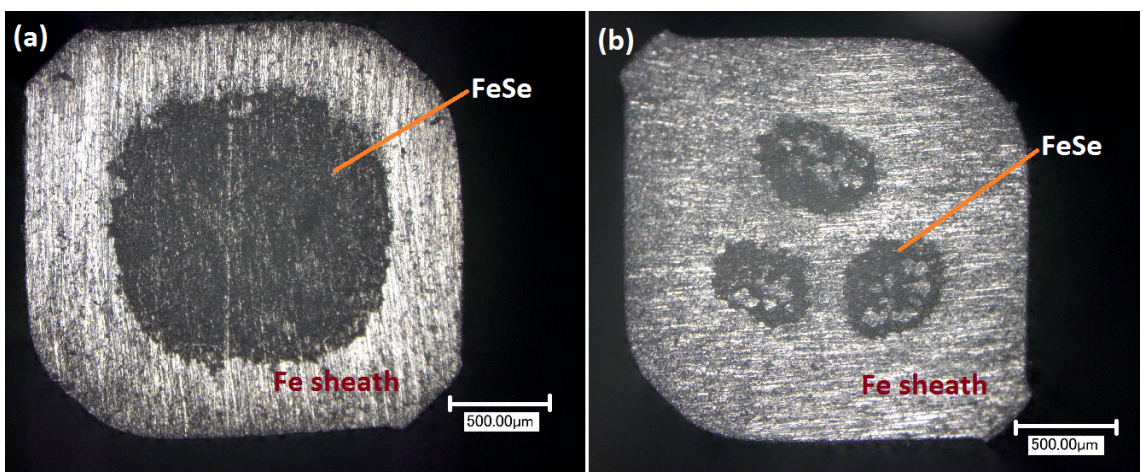

Fig. 3

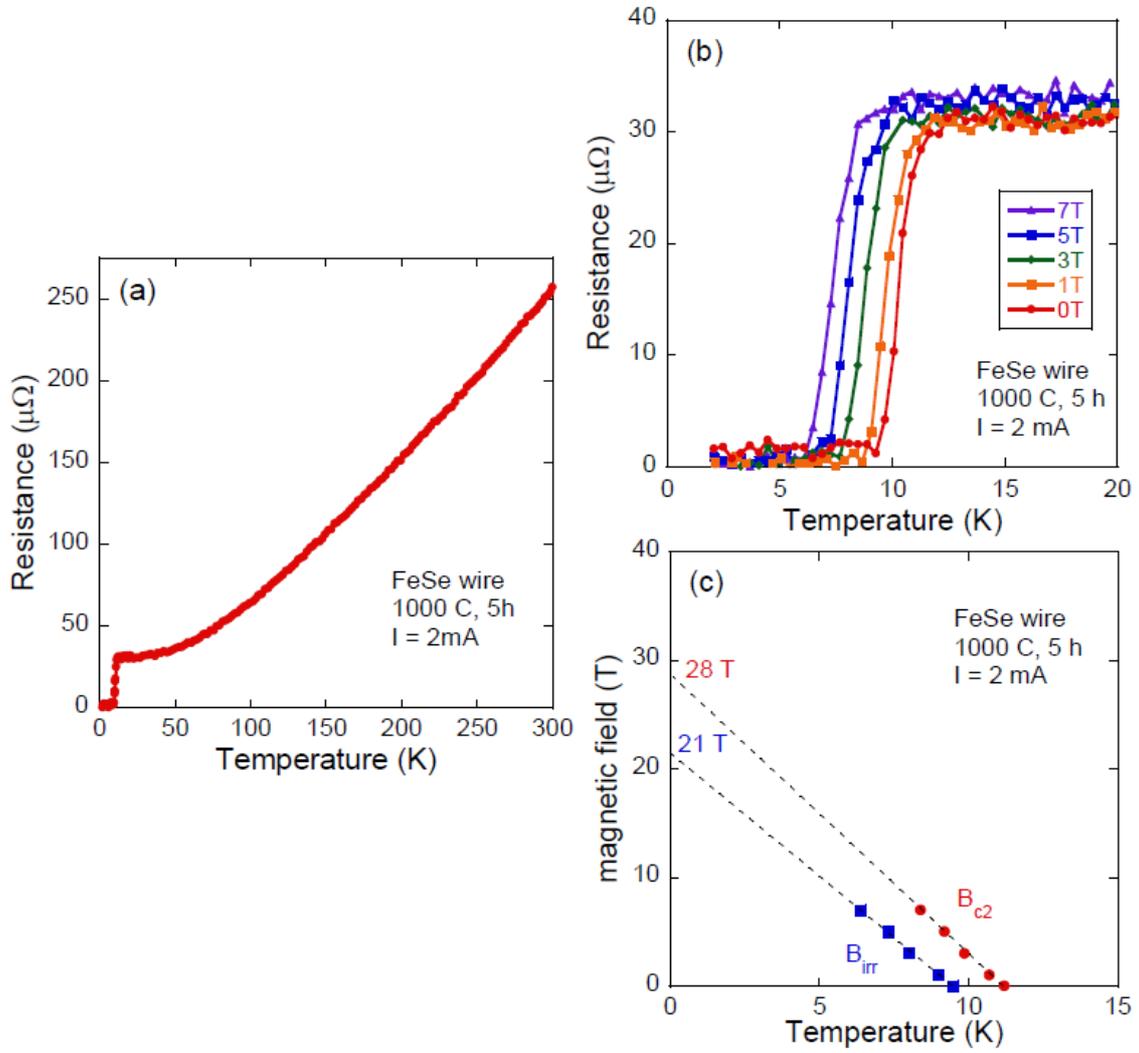

Fig. 4

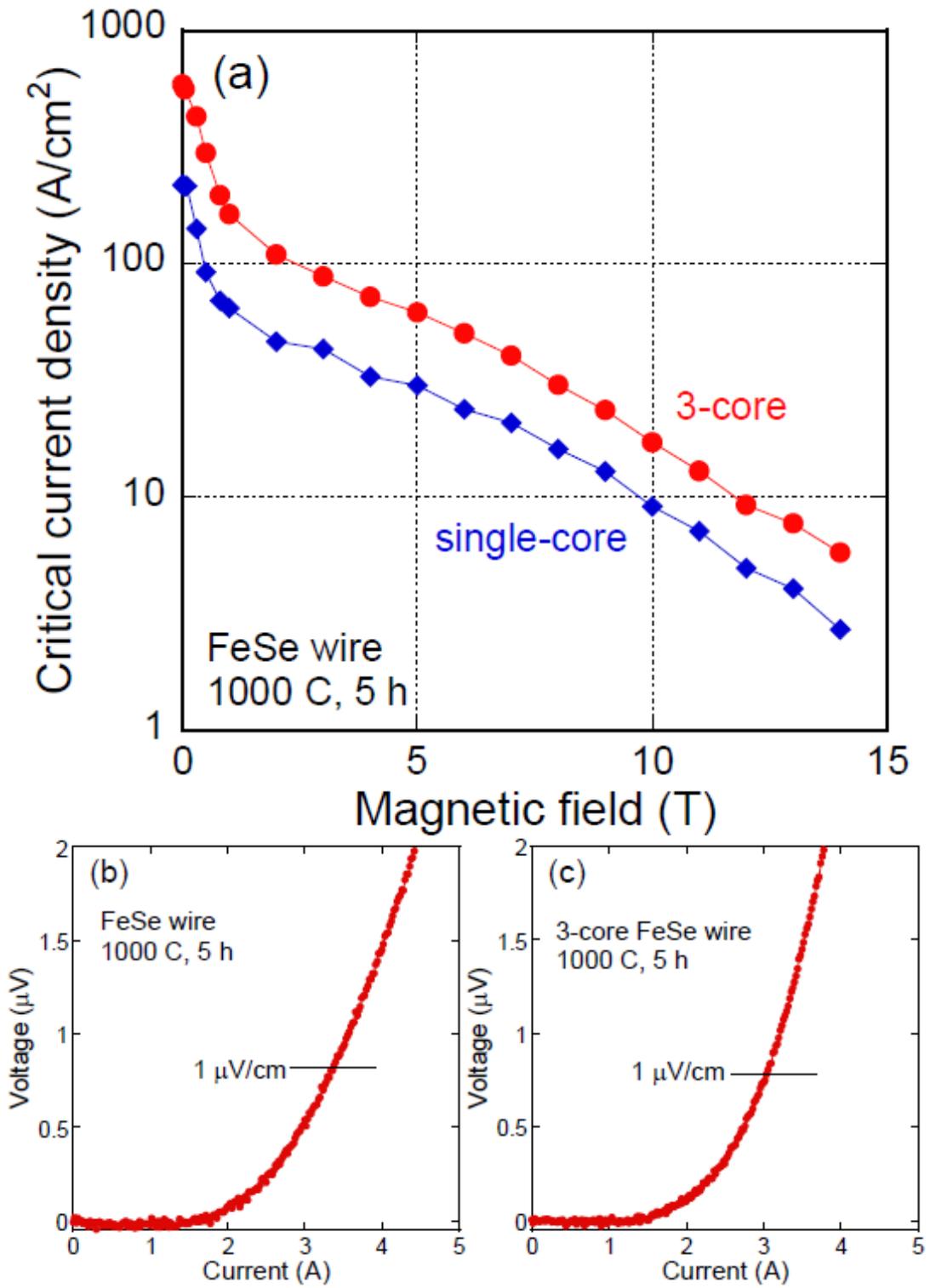